%% file: Template.tex
\documentclass{article}
\usepackage{spconf,graphicx,hyperref}
\ninept
\usepackage{algorithm}
\usepackage{algpseudocode}
\usepackage{amsfonts}

\usepackage{colortbl}

\usepackage{cite}
\usepackage{amsmath,amssymb,amsfonts}
\usepackage{textcomp}
\usepackage{xcolor}
\usepackage{multirow}
\usepackage{soul}
\newcommand{\swap}[1]{{\color{black}#1}}
\newcommand{\sanjeel}[1]{{\color{black}#1}}

\newcommand\ie{\textit{i.e.}}
\newcommand{\parlabel}[1]{\vspace{0.5em}\noindent\textbf{#1}.}

\definecolor{Mycolor2}{HTML}{ccffff}
\definecolor{Mycolor3}{HTML}{dbdfde}

\newcommand{\ours}{\cellcolor{Mycolor3}}

\newcommand{\metriccol}{\cellcolor{Mycolor2}}
\def\modelname{\texttt{HypEE}}
\def\baseline{\texttt{EucEE}}

\usepackage{booktabs}

\title{More Than a Shortcut: A Hyperbolic Approach to Early-Exit Networks}
\name{\begin{tabular}{c}
      Swapnil Bhosale$^{1^\dagger}$\thanks{${\dagger}$ Work done during internship at Meta Reality Labs Research, UK.}, 
      Cosmin Fr{\u{a}}teanu$^{2}$, 
      Camilla Clark$^{2}$, 
      Arnoldas Jasonas$^{2}$, 
      Chris Mitchell$^{2}$, \\
      Xiatian Zhu$^{1}$, 
      Vamsi Krishna Ithapu$^{2}$, 
      Giacomo Ferroni$^{2}$,
      {\c{C}}a{\u{g}}da{\c{s}} Bilen$^{2}$, 
      Sanjeel Parekh$^{2}$
      \end{tabular}}

\address{$^{1}$University of Surrey, UK. \qquad $^{2}$Meta Reality Labs Research, UK.}

\begin{document}

\maketitle
\input{sec/0_abstract}    
\input{sec/1_intro}
\input{sec/2_related_work}

\input{sec/3_method}
\input{sec/4_experiments}

\input{sec/supple_triggering}
\input{sec/5_conclusion}

\appendix
\input{sec/supple_motivation_delta}
\input{sec/supple_early_exit_analysis}
\input{sec/supple_additional_vis}
\input{sec/supple_lookahead}
\input{sec/supple_traversal}

\bibliographystyle{IEEEbib}
\bibliography{strings,refs}

\end{document}

%% file: sec/0_abstract.tex
\begin{abstract}
Deploying accurate event detection on resource-constrained devices is challenged by the trade-off between performance and computational cost. 
While Early-Exit (EE) networks offer a solution through adaptive computation, they often fail to enforce a coherent hierarchical structure, limiting the reliability of their early predictions.
To address this, we propose Hyperbolic Early-Exit networks (\modelname{}), a novel framework that learns EE representations in hyperbolic space. Our core contribution is a hierarchical training objective with a novel entailment loss, which enforces a partial-ordering constraint to ensure that deeper network layers geometrically refine the representations of shallower ones. 
Experiments on multiple audio event detection tasks and backbone architectures show that \modelname{} significantly outperforms standard Euclidean EE baselines, especially at the earliest, most computationally-critical exits. 
The learned geometry also provides a principled measure of uncertainty, enabling a novel triggering mechanism that makes the overall system both more efficient and more accurate than a conventional EE and standard backbone models without early-exits.

\end{abstract}


%% file: sec/1_intro.tex
\section{Introduction}
\label{intro}
The proliferation of ``always-on'' \swap{audio} sensing applications on resource-constrained wearable devices, from health monitoring to safety alerts, has created a pressing need for highly efficient event detection systems \cite{dorschky2015framework, shankar2024edge}.
These applications necessitate a careful balance between computational efficiency and detection accuracy, owing to stringent limitations on power consumption, memory, and real-time response requirements \cite{Prince2019DeployingAcoustic}. 
At the core of this problem lies a fundamental trade-off: lightweight, low-compute detectors offer rapid, energy-efficient predictions but often lack the robustness required for diverse and unpredictable acoustic environments such as high background noise or overlapping audio events. Conversely, complex, high-accuracy models are too power-hungry for the continuous operation mandated by wearable use cases.

\begin{figure}[t]
    \centering
    \resizebox{0.85\columnwidth}{!}{%
        \includegraphics{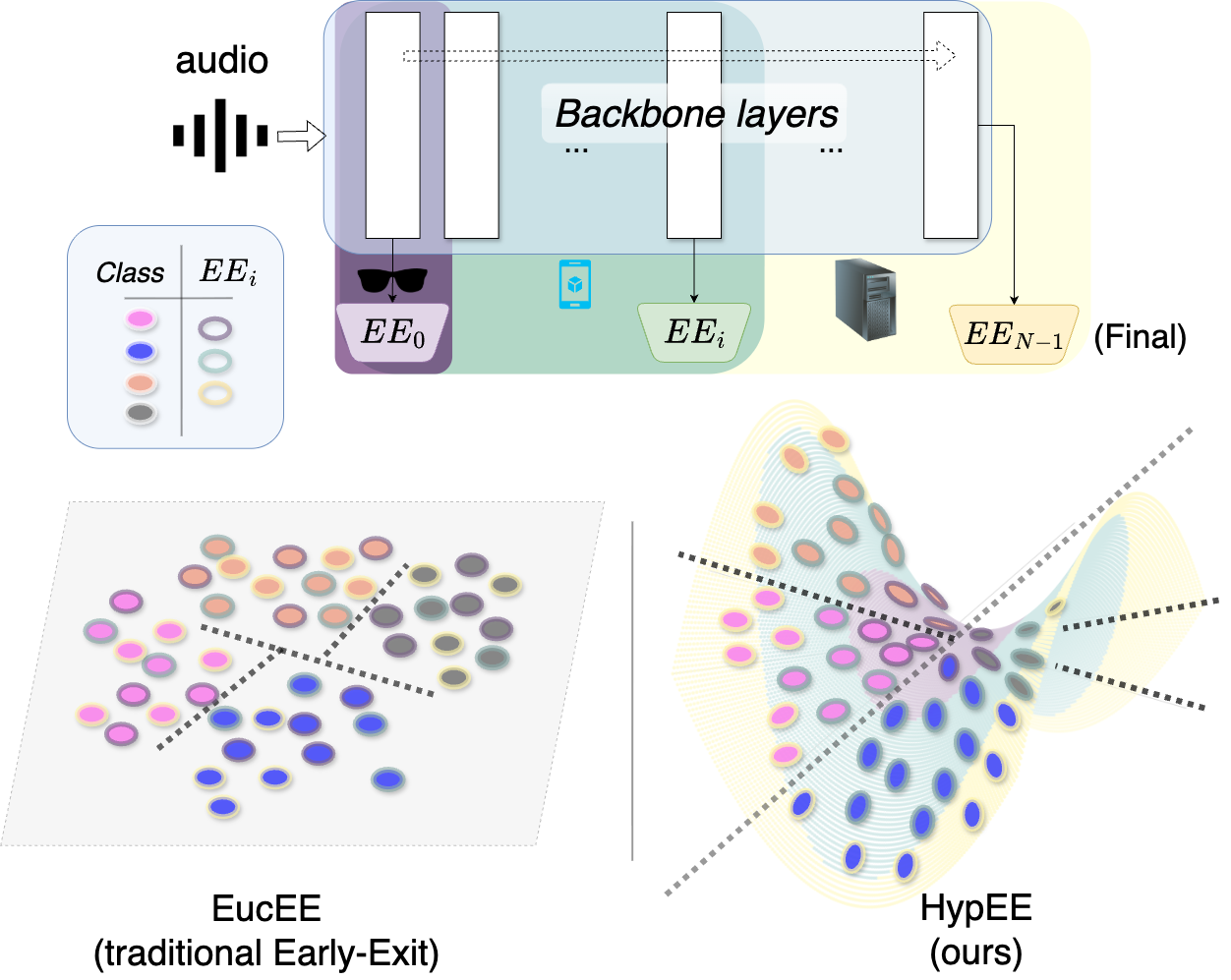}
    }
    \vspace{-1em}
    \caption{Our multi-stage system (\textbf{Top}) deploys early-exits on devices with varying resources, from glasses (${EE}_{0}$) to servers. 
    While a standard Euclidean approach (\textbf{Bottom-left}) fails to learn a structured latent space, our hyperbolic model, \modelname{} (\textbf{Bottom-right}), learns a meaningful hierarchy, separating classes angularly and exit levels radially based on certainty.
    }
    \label{fig:teaser}
    \vspace{-1.8em}
\end{figure}

\swap{As shown in Fig. \ref{fig:teaser} (Top), Early-Exit (EE) networks provide a natural architectural framework for balancing this trade-off through a multi-stage approach: lightweight exits efficiently handle common or easy inputs, while difficult or uncertain inputs progress to deeper, more complex stages for refined analysis.}
\cite{panda2016conditional}. This approach has shown promise for optimizing the trade-off between efficiency and performance \cite{Zhang2021BasisNet}, especially in scenarios where the difficulty of event detection can vary widely across input instances.
However, the practical efficacy of EE networks hinges on two critical challenges that are often inadequately addressed. 
First, \swap{traditional EE training fails} to enforce a coherent, hierarchical relationship between the exits \cite{dorschky2015framework, rahmath2024early}, causing them to behave as independent classifiers rather than a sequence of refining stages. This lack of partial-ordering \cite{Vendrov2015OrderEmbeddings} undermines the reliability of early \sanjeel{stages, leading to potentially conflicting predictions}. 
Second, the decision to exit is typically guided by heuristics like softmax confidence, which are known to be poorly calibrated and unreliable measures of a model's true uncertainty \cite{nguyen2015deep, guo2017calibration}.

\textbf{Our Contributions}. To address these fundamental issues, we propose \modelname{}, a novel framework that reframes the Early Exit paradigm by explicitly modeling the hierarchy inherent in a multi-stage system. In \modelname{}, each stage progressively refines the representation of its predecessor, becoming increasingly certain about the classification output. We formalize this process using hyperbolic geometry which naturally lends itself to such hierarchical learning and uncertainty-aware inference \cite{nickel2017poincare, mathieu2019continuous}. In particular, we introduce two key methodological contributions: (1) A \textbf{hierarchical training objective} with an \textit{entailment loss} that uses adaptive geometric cones to ensure that deeper network layers systematically refine the representations of shallower ones, 
(2) A \textbf{geometry-aware triggering algorithm} that operationalizes the learned structure by using the distance of an embedding from the origin of the hyperboloid as a direct and robust measure of model uncertainty. This provides a more reliable way than conventional entropy-based heuristics to determine whether or not to trigger subsequent stages of compute.
As illustrated in Fig. \ref{fig:teaser} (Bottom), while traditional Euclidean latent spaces fail to capture the partial-ordering of the exits, \modelname{} learns a representation that simultaneously organizes samples by class (angularly) and by exit-level (radially). The radial structure, where proximity to the origin corresponds to higher uncertainty \cite{ganeaentailmentcone}, is a direct result of our entailment objective and provides a principled foundation for adaptive computation in multi-stage systems.
We demonstrate the effectiveness of our approach through extensive experiments across 
two audio recognition tasks (audio tagging and event detection), and 
two backbone architectures (Transformer-based and CNN-based), showing significant performance and efficiency gains, especially for low-compute early exits.

%% file: sec/2_related_work.tex
\section{Related works}

\parlabel{Hyperbolic Geometry in Deep Learning} Recent research has highlighted advantages of using hyperbolic geometry as a prior for the feature space of neural networks \cite{mettes2024hyperbolic, desai2023hyperbolic, pal2024compositional}. Unlike Euclidean space, hyperbolic spaces can be conceptualized as continuous versions of trees, making them naturally suited for embedding hierarchical or taxonomic data with minimal distortion. This capacity stems from the property that the volume of a hyperbolic ball grows exponentially with its radius, allowing it to efficiently accommodate tree-like structures \cite{nickel2017poincare, sala2018representation, tifrea2018poincar}. Consequently, Hyperbolic Neural Networks (HNNs) \cite{peng2021hyperbolic, bdeir2023fully} have demonstrated superior performance in domains with inherent hierarchies, such as natural language processing \cite{he2025hyperbolic}, graph analytics \cite{liu2019hyperbolic} and vision \cite{mettes2024hyperbolic}. This motivation has also extended to the audio domain, with a focus on tasks such as hierarchical source separation \cite{petermann2023hyperbolic, petermann2024hyperbolic}, anomaly detection \cite{germain2023hyperbolic} and multimodal learning \cite{hong2023hyperbolic}.
Our goal is different -- we observe that  features in a neural network are organized hierarchically, from simple patterns in early layers to complex abstractions in deeper ones. 
This inherent structure makes hyperbolic geometry ideal for modeling the progressive refinement required in Early-Exit networks.

\parlabel{Hierarchical Audio Classification} Existing literature explores hierarchical classification in audio by \emph{imposing} pre-defined taxonomies based on 
biology \cite{cramer2020chirping}, 
acoustic scenes \cite{khandelwal2023multi}, 
or sound types \cite{liang2024learning}
—requiring domain expertise and fixed structural assumptions.
In contrast, our work fundamentally shifts the objective from, 
only classifying what a sound is to navigating the dynamic stages of a model's own inference process. 
Particularly, \modelname{} uses hyperbolic geometry to learn an emergent hierarchy of a model's intermediate representations, rather than only semantics. 
This makes it uniquely suited to enforce progressive refinement in multi-stage systems, a task for which static taxonomies are ill-suited.

%% file: sec/3_method.tex
\section{\modelname{}: Hyperbolic Early-Exit Networks}

We build \modelname{} upon the \emph{Lorentz} model of hyperbolic geometry, chosen for its numerical stability over alternatives like the \emph{Poincaré} ball model \cite{Mishne2023NumericalHyperbolic}. 
This model represents an $n$-dimensional hyperbolic space on the upper sheet of a two-sheeted hyperboloid embedded in an $(n+1)$-dimensional space \cite{Minkowski1908RaumZeit}.
Following conventions from special relativity \cite{Einstein2015PrincipleRelativity}, this ambient space is described with one \textit{time} dimension and $n$ \textit{space} dimensions.

\begin{figure}[t]
    \centering
    \resizebox{\columnwidth}{!}{%
        \includegraphics{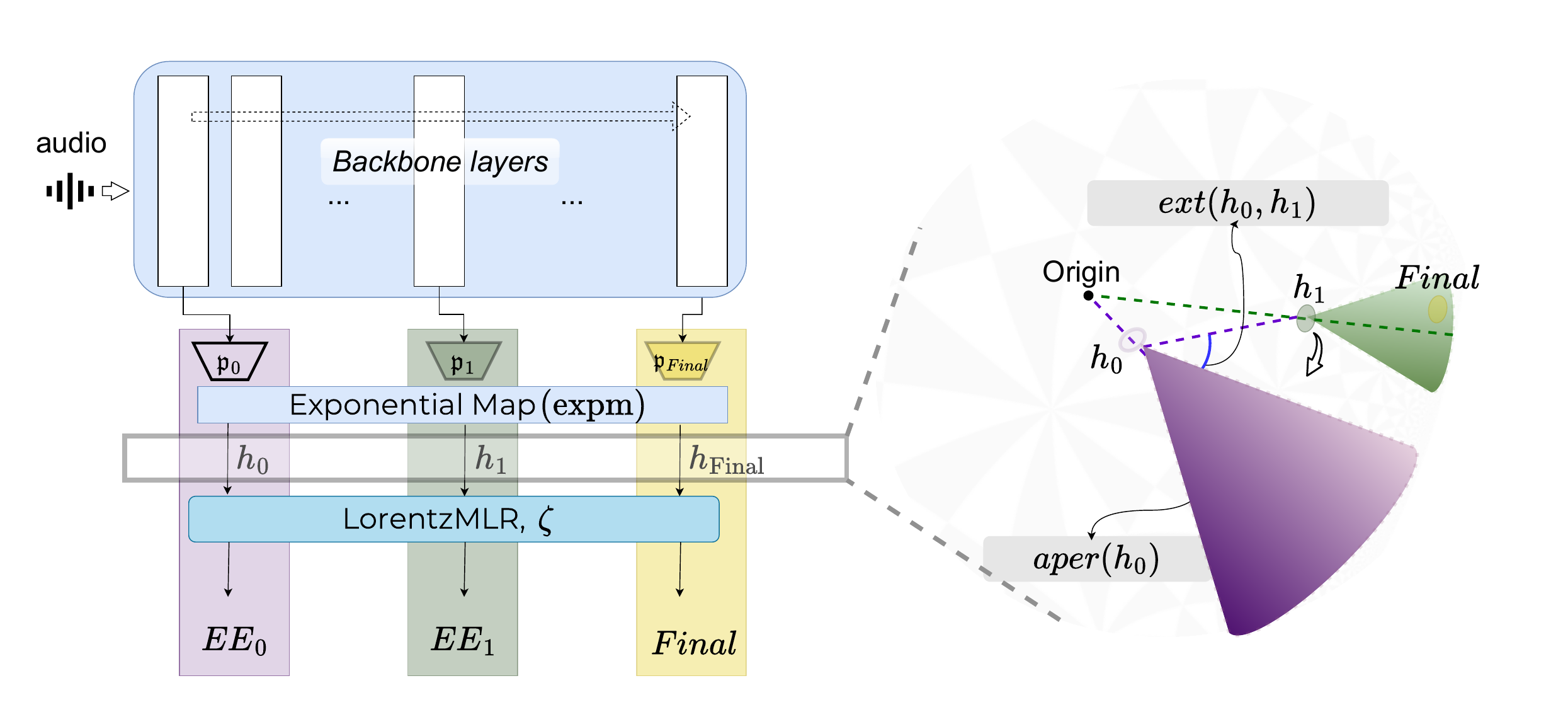}
    }
    \vspace{-2.5em}
    
    \caption{The \modelname{} framework. \textbf{Left:} Euclidean embeddings are mapped to the \emph{Lorentz} hyperboloid, and a hierarchical entailment loss enforces a partial-order constraint on embeddings from consecutive exits. \textbf{Right:} In the resulting latent space, 
    \modelname{} learns to organize embeddings radially by exit-level and angularly by class, forming trajectories that move outwards as certainty increases, whilst forcing entailment across successive exits (see arrow direction).}
    \label{fig:method}
    \vspace{-1.4em}
\end{figure}

A point $x$ in this $(n+1)$-dimensional space lies on the hyperboloid if it satisfies $\langle x, x \rangle_\mathcal{L} = -1/c$, where $c>0$ is a constant related to the space's curvature and $\langle \cdot, \cdot \rangle_\mathcal{L}$ is the Lorentzian inner product \cite{Law2019Lorentzian}. The distance between two points on this curved surface is measured by the geodesic distance, $d_\mathcal{L}(x, y) = \frac{1}{\sqrt{c}} \cosh^{-1}\left( -c \langle x, y \rangle_\mathcal{L} \right)$.
Our architecture, depicted in Fig. \ref{fig:method}, begins by taking a standard Euclidean embedding, $z_{i} \in \mathbb{R}^{n}$, 

\swap{\ie{} the output at the corresponding intermediate layer of the backbone where the Early-Exit $i$ is placed (post projection $\mathfrak{p}_{i}(.)$ to match dimensions across all exits.}
To create a hierarchical representation, we map $z_i$ (Euclidean vector) onto the curved \emph{Lorentz} hyperboloid. 
Specifically, we treat $z_i$ as a vector in the tangent space at the origin and project it using the exponential map ($\sanjeel{\text{expm}}$) \cite{Nickel2018LorentzHyperbolic}, yielding the final hyperbolic embedding: 
$h_i (\in \mathbb{L}_{c}^{n}) = \text{expm}_{\mathbf{o}}([z_i, 0])$.
Preserving numerical stability, we scale the Euclidean vectors with learnable scalars \cite{desai2023hyperbolic} before projection.
Classification is then handled in hyperbolic space by a \emph{Lorentz} Multinomial Logistic Regression (MLR) classifier \cite{bdeir2023fully}\swap{, $\zeta(.)$}, which calculates logits from the signed hyperbolic distance of the embedding $h_i$ to a set of class-defining hyperplanes.

\subsection{Hierarchical Training with Entailment Loss}
To ensure that the exits act as a sequence of refinement stages, we design a training objective that combines a standard cross-entropy loss with a hierarchical consistency loss. The total loss for the network is a weighted sum:

\sanjeel{
$$\mathcal{L}_{\text{total}}=\sum_{\swap{i=0}}^{N-1}w_{i} \cdot \mathcal{L}_{\text{class}}(\swap{\zeta(h_{i})},y)+\lambda \cdot \sum_{\swap{i=0}}^{N-2}\mathcal{L}_{\text{entail}}(h_{i+1},h_{i})$$
}
Here, 
\sanjeel{$\mathcal{L}_{\text{class}}$} is the standard cross-entropy classification loss applied to the logits from the Lorentz MLR at each exit \sanjeel{$i$}, weighted by $w_{i}$. The second term, 
\sanjeel{$\mathcal{L}_{\text{entail}}$}, is an 
entailment loss that enforces a partial-order constraint between the hyperbolic embeddings of consecutive exits, $h_{i}$ and $h_{i+1}$.
Inspired by work on learning concept hierarchies \cite{desai2023hyperbolic}, we define an 
entailment cone for each shallow embedding $h_{i}$. The loss is designed to constrain the deeper embedding 
$h_{i+1}$ to lie within this cone, mathematically ensuring that the prediction at stage $i$ \textit{entails} the more specific one at stage $i+1$. The loss for a pair of consecutive exits is formulated as:
$$\sanjeel{\mathcal{L}_{\text{entail}}}(h_{i+1},h_{i})=\max(0,ext(h_{i},h_{i+1})-aper(h_{i})),$$
where 
$ext(h_{i},h_{i+1})$ is the exterior angle between the origin, $h_{i}$, and $h_{i+1}$, and $aper(h_{i})$ is the half-aperture (width) of the cone projected by $h_{i}$, shown in Fig.~\ref{fig:method} (Right).
Crucially, this formulation provides an adaptive mechanism for learning since the aperture of the cone, 
$aper(h_{i})$, 
is defined to be inversely proportional to the certainty of the embedding $h_{i}$, 
which in the \emph{Lorentz} model corresponds to its distance from the origin.
If an EE embedding 
$h_{i}$ is uncertain (close to the origin), 
its entailment cone is wide, 
granting the next layer $h_{i+1}$ more freedom to adjust and refine the representation.
Conversely, if 
$h_{i}$ is certain (far from the origin), 
its cone is narrow, 
enforcing consistency and preventing the deeper layers from drastically altering an already confident prediction.
This geometry-aware objective elegantly models the desired \textit{consistency-then-refinement} dynamic across the network's depth, directly addressing the lack of partial-ordering in traditional Early-Exit models.

%% file: sec/4_experiments.tex
\section{Experiments}

We demonstrate the effectiveness of our proposed \modelname{} network across multiple audio tasks and backbone architectures. \modelname{} is compared with a strong Euclidean Early-Exit baseline (\baseline{}).

\noindent\textbf{Backbone Architectures.} We evaluate our approach on two distinct backbones to test its versatility: \textbf{BEATs} \cite{chen2022beats}, a state-of-the-art Transformer-based audio model, and \textbf{MobileNetV3} \cite{koonce2021mobilenetv3}, a lightweight and efficient convolutional neural network. For the BEATs backbone, the exits are placed after layers 1 (${EE}_{0}$), 3 (${EE}_{1}$), and the final layer 12, corresponding to computational costs of 0.24M, 0.71M, and 2.86M MMACs, and parameter counts of 11.81M, 25.99M, and 90.61M, respectively. Similarly, for MobileNetV3, the exits are placed after layers 8 (${EE}_{0}$), 12 (${EE}_{1}$), and the final layer 17, with costs of 13.08K, 19.41K, and 34.9K MMACs, and parameter counts of 0.077M, 0.42M, and 3.1M, respectively.

\noindent\textbf{Baseline.} Our baseline, a standard Euclidean Early-Exit (\baseline{}) network, uses the same backbone architectures and exit placements as \modelname{}. However, its exits consist of standard linear classifiers, and the entire model is trained and operates within a Euclidean latent space. For a fair and strong comparison, \baseline{} is trained using the ``mixed'' training strategy \cite{kubatytrain}, which is shown to be highly effective for conventional EE models.

\noindent\textbf{\swap{Tasks, Datasets and Metrics}.} 
We assess performance across two primary audio tasks. For \textsc{Audio Tagging (AT)}, we augment the widely used \textbf{ESC-50} \cite{piczak2015esc} dataset with soundbanks from UrbanSound8K \cite{Salamon2014UrbanSound} dataset creating a version five times larger than the original. 
We report classification accuracy across 50 audio event tags using 5-fold cross-validation. 
For \textsc{Sound Event Detection (SED)}, we evaluate on 
the large-scale, real-world \textbf{Audioset Strong} \cite{hershey2021benefit} dataset (407 classes). 
Performance for the SED task is measured using the macro-averaged Polyphonic Sound Detection Score (PSDS) \cite{bilen2020framework} and the Area Under the ROC Curve (\textbf{AUROC}).

\subsection{Effectiveness of the Hierarchical Training Objective}

\subsubsection{Quantitative Analysis}
\noindent\textbf{Audio Tagging.} In Table \ref{tab:all_combined} (rows 1-2), \modelname{} demonstrates a significant improvement over the Euclidean baseline, particularly at the earliest and most computationally constrained exit, ${EE}_{0}$. With the BEATs backbone, \modelname{} boosts the accuracy at ${EE}_{0}$ from 58.32\% to 82.19\%, an absolute improvement of over 23\%. 
Similarly, on MobileNetV3, accuracy at ${EE}_{0}$ jumps from 43.32\% to 62.08\%. 
The gains at the initial exit highlight the ability of our geometrically structured approach to produce highly reliable predictions with minimal computation. The performance advantage is maintained across the deeper exits as well.

\begin{table}[t]
\vspace{-7pt}
\centering
\caption{
Comparing \modelname{} with Euclidean baseline, \baseline{} for the Audio Tagging and Sound Event Detection task.
}
\label{tab:all_combined}
\resizebox{\columnwidth}{!}{%
\begin{tabular}{c|c|c|cc|cc|cc}
\toprule
& \textbf{Backbone} & \textbf{Method} 
& \multicolumn{2}{c|}{${EE}_{0}$}
& \multicolumn{2}{c|}{${EE}_{1}$}
& \multicolumn{2}{c}{Final} \\

\midrule\midrule

 \multicolumn{3}{c|}{\textcolor{blue}{Audio Tagging}}
& \multicolumn{2}{c|}{\metriccol\textbf{Accuracy}} 
& \multicolumn{2}{c|}{\metriccol\textbf{Accuracy}} 
& \multicolumn{2}{c}{\metriccol\textbf{Accuracy}} \\

\midrule
\parbox[t]{3mm}{\multirow{4}{*}{\rotatebox[origin=c]{90}{\footnotesize ESC-50}}} & \multirow{2}{*}{BEATs} & \baseline{}
& \multicolumn{2}{c|}{58.32} 
& \multicolumn{2}{c|}{83.42} 
& \multicolumn{2}{c}{92.14} \\
&  & \ours\modelname{}
& \multicolumn{2}{c|}{\ours\textbf{82.19}} 
& \multicolumn{2}{c|}{\ours\textbf{90.01}} 
& \multicolumn{2}{c}{\ours\textbf{93.16}} \\
\cmidrule(lr){2-9}
 & \multirow{2}{*}{MobileNetV3} & \baseline{}
& \multicolumn{2}{c|}{43.32} 
& \multicolumn{2}{c|}{62.57} 
& \multicolumn{2}{c}{81.32} \\
&  & \ours\modelname{}
& \multicolumn{2}{c|}{\ours\textbf{62.08}} 
& \multicolumn{2}{c|}{\ours\textbf{71.32}} 
& \multicolumn{2}{c}{\ours\textbf{83.39}} \\

\midrule\midrule

 \multicolumn{3}{c|}{\textcolor{blue}{Sound Event Detection}}  & \metriccol\textbf{PSDS} & \metriccol\textbf{AUROC} & \metriccol\textbf{PSDS} & \metriccol\textbf{AUROC} & \metriccol\textbf{PSDS} & \metriccol\textbf{AUROC} \\
\midrule

\parbox[t]{3mm}{\multirow{4}{*}{\rotatebox[origin=c]{90}{\footnotesize Audioset-S}}} & \multirow{2}{*}{BEATs} & \baseline{} & 9.25  & 40.68 & 25.24 & 58.35 & \textbf{44.80} & \textbf{82.75} \\
& & \ours\modelname{} & \ours\textbf{16.97} & \ours\textbf{46.47} & \ours\textbf{32.26} & \ours\textbf{67.29} & \ours43.59 & \ours80.48 \\
\cmidrule(lr){2-9}
& \multirow{2}{*}{MobileNetV3} & \baseline{} & 12.30 & 45.66 & 9.42  & 38.48 & \textbf{39.93} & \textbf{76.11} \\
& & \ours\modelname{} & \ours\textbf{18.71} & \ours\textbf{50.74} & \ours\textbf{23.47} & \ours\textbf{54.87} & \ours38.12 & \ours72.75 \\

\bottomrule
\end{tabular}
}
\vspace{-7pt}
\end{table}

\noindent\textbf{Sound Event Detection.} 
The results for SED, shown in Table \ref{tab:all_combined} (rows 3-4)
for Audioset Strong,
confirm a similar trend. 
Across both the backbones, \modelname{} consistently outperforms the baseline, especially at the early-exits ${EE}_{0}$ and ${EE}_{1}$. 
With the BEATs backbone, \modelname{} improves the macro-averaged PSDS at ${EE}_{1}$ from 25.24 to 32.26. 
Albeit, at the final exit, the performance on both backbones converges (since both utilize the full capacity of the backbone network), the primary benefit of \modelname{} is its ability to drastically improve the quality of early, low-cost predictions.

\begin{figure}[t]
    \centering
    \resizebox{\columnwidth}{!}{%
        \includegraphics{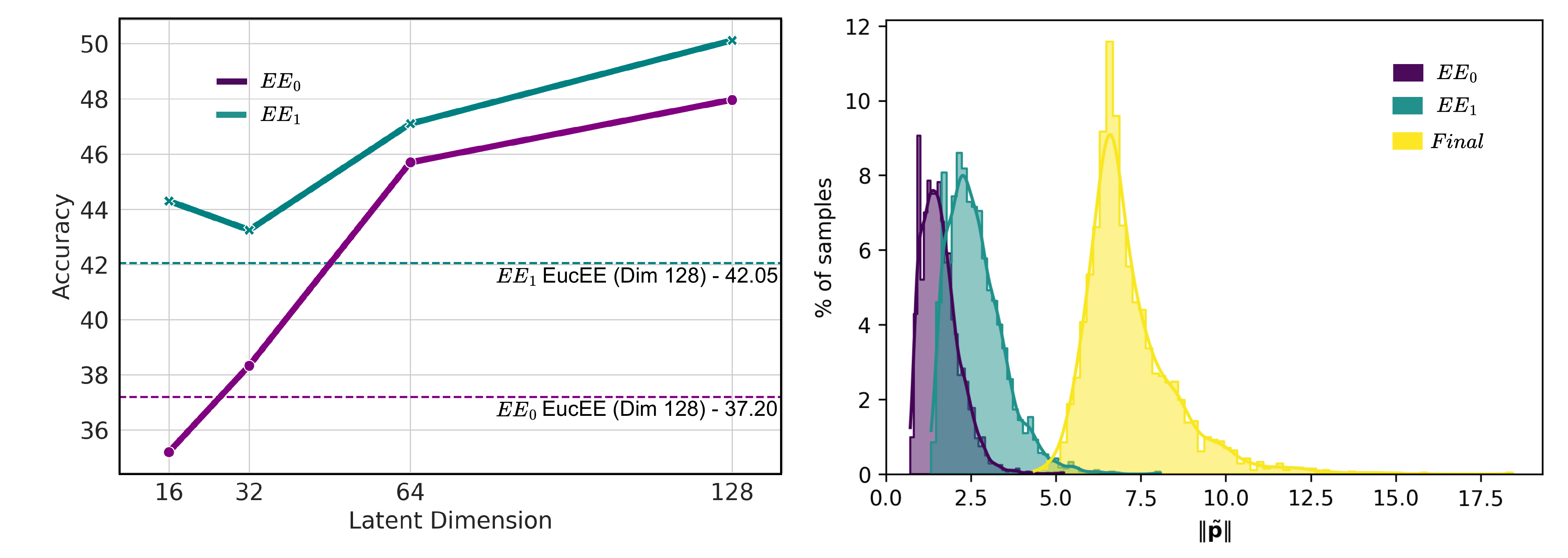}
    }
    \vspace{-1.7em}
        \caption{\textbf{Left:} Effect of Latent Dimension on Early-Exit Performance (${EE}_{0}$, ${EE}_{1}$) for \baseline{} and \modelname{}. \textbf{Right:} Distribution of embedding norms $\|\tilde{p}\|$ for each exit, showing a clear separation and ordering, where earlier exits (${EE}_{0}$) are closer to the origin, indicating a learned hierarchy of refinement.}
    \label{fig:latent_dim_combined_hist}
    \vspace{-1.8em}
\end{figure}

\noindent\textbf{Embedding Space Efficiency.} We hypothesize that hyperbolic geometry uses embedding space more efficiently. An ablation study on the latent dimension size, $n$ (Fig. \ref{fig:latent_dim_combined_hist}, Left), for the AT task, confirms this: \modelname{} with as low as $n = 32$, a 4x reduction achieves performance comparable to the 128-dimension Euclidean baseline. Overall, \modelname{} learns powerful and compact representations, making it well-suited for memory-constrained wearable devices.

\subsubsection{Qualitative Analysis}
\label{fig:qualitative_main_paper}

\begin{figure}[t]
    \centering
    \resizebox{\columnwidth}{!}{%
        \includegraphics{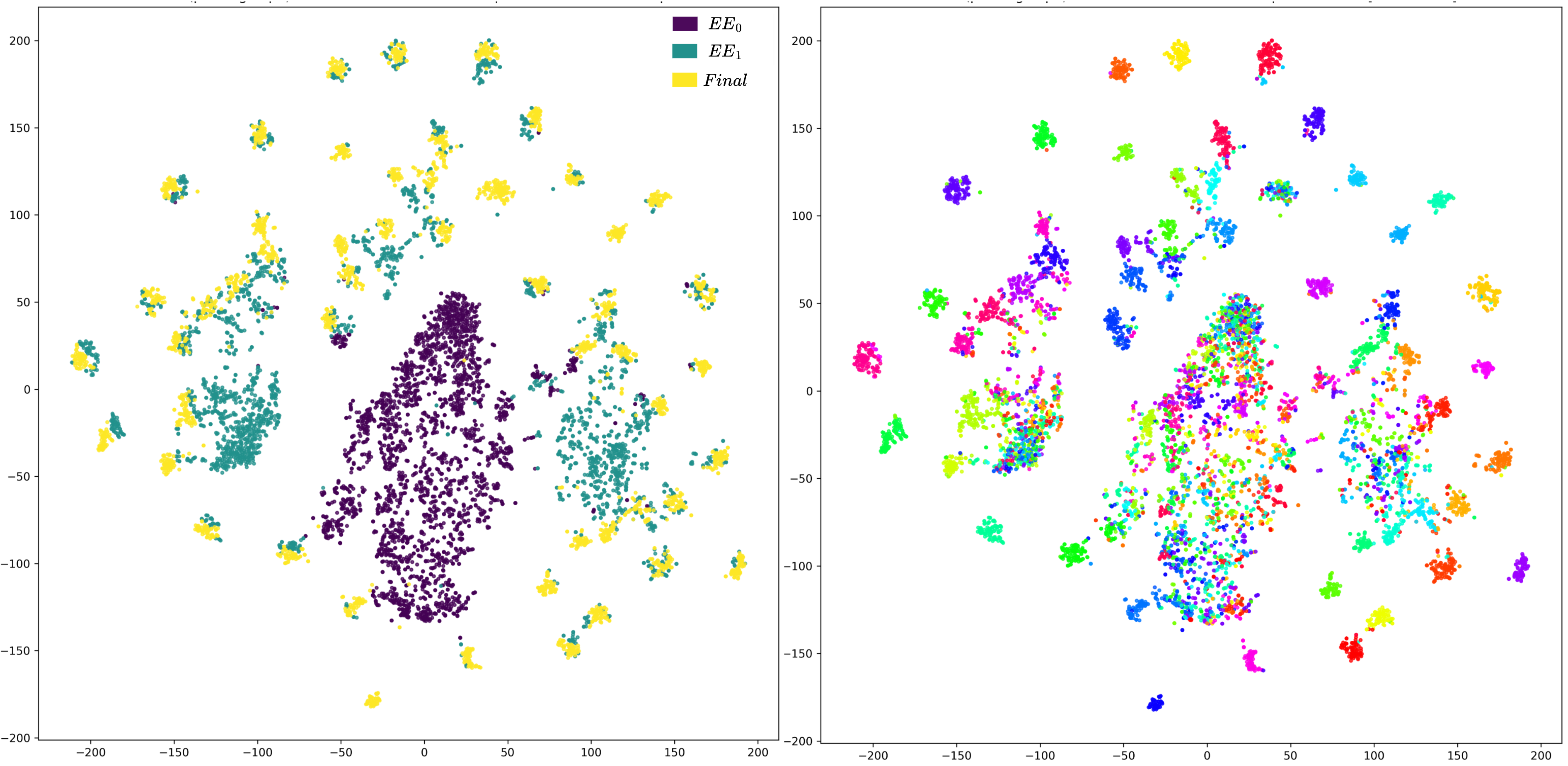}
    }
    \vspace{-1.5em}
    
    \caption{t-SNE of hyperbolic embeddings in the tangent space, confirming a dually-structured latent space. \textbf{Left:} Coloring by exit level reveals a clear hierarchy, with early exits (${EE}_{0}$) forming a core refined by later ones. \textbf{Right:} Coloring by class label shows strong semantic clustering.}
    \label{fig:logmap_tsne}
    \vspace{-1.8em}
\end{figure}

We verify that our entailment loss organizes the latent space into a hierarchy by visualizing the embeddings from the ESC-50 experiment in two ways. 
First, inspired by recent work in compositional hyperbolics \cite{pal2024compositional}, we analyze the distribution of the spatial norms of the hyperbolic embeddings, $\|\tilde{p}\|$, which corresponds to their distance from the origin. 
Fig. \ref{fig:latent_dim_combined_hist}, Right, plots these distributions for embeddings from each of \modelname{}'s exits (${EE}_{0}$, ${EE}_{1}$, and $Final$). 
The embeddings are clearly separated by exit;
those from the first exit (${EE}_{0}$) are tightly clustered closest to the origin, followed by the second exit (${EE}_{1}$), with the final exit's embeddings pushed furthest away. This visually confirms the intended hierarchy, positioning earlier, more uncertain representations closer to the ``root'', while more refined and certain representations from deeper layers are pushed outwards.
Second, we project the hyperbolic embeddings to 2D Euclidean space for visualization using a log-map and t-SNE (Fig. \ref{fig:logmap_tsne}). Coloring by exit level (Left) reveals a structural hierarchy where early exits form a core that is refined by later ones thus, demonstrating a \textit{consistency-then-refinement} dynamic. 
Coloring by ground-truth class (Right) shows distinct semantic clusters. Taken together, this demonstrates that \modelname{} learns a latent space simultaneously structured by both the exit hierarchy and the class semantics.

%% file: sec/supple_triggering.tex
\subsection{Uncertainty-Gated Triggering with \modelname{}}
\label{sec:trigger}

To recall, the goal of a triggering mechanism is to automatically determine the best early-exit for a given input. This is typically done by utilizing an estimate of the model's confidence at each exit.
A key advantage of the hierarchically structured space learned by \modelname{} is that the geometry itself provides a robust, principled measure of model uncertainty. While standard EE models typically rely on heuristics like softmax confidence or entropy for their triggering criteria—which are often poorly calibrated—we can instead directly use the distance of an embedding from the origin of the hyperboloid. 

To fully harness this property, we demostrate a proof-of-concept, geometry-aware triggering algorithm for inference, detailed in Algorithm 1.
\begin{algorithm}
\label{alg:triggering_criteria}
{\footnotesize
\caption{Uncertainty-Gated Triggering with \modelname{}}
\begin{algorithmic}

\Require\\
\begin{itemize}
\setlength\itemsep{-0.1em}
    \item Pre-trained backbone with $L$ layers, EE gates at $N<L$ layers.
    \item Global mean ($\mu_{i}^{\text{correct}}, \mu_{i}^{\text{incorrect}}$) and std ($\sigma_{m}^{\text{correct}}, \sigma_{m}^{\text{incorrect}}$) of hyperbolic norm and confidence for each exit $i$.
    \item Per-class mean ($\mu_{i,c}^{\text{correct}}, \mu_{i,c}^{\text{incorrect}}$) and std ($\sigma_{i,c}^{\text{correct}}, \sigma_{i,c}^{\text{incorrect}}$) for every gate $i$ and class $c$.
    \item Query batch $\mathcal{X} = \{x^{(j)}\}_{j=1}^M$ for Early-Exit inference.
\end{itemize}

\For{each exit gate $EE_i$ in order}
    \State Compute embedding $e_i$ at exit $EE_i$ for input $x$
    \State Compute embedding norm $\|e_i\|$
    
    \State $p_{\text{correct}} \gets \mathcal{N}(\|e_i\|; \mu^{\text{correct}}_i, \sigma^{\text{correct}}_i)$ \Comment{\textcolor{cyan}{Prob. under correct dist.}}
    \State $p_{\text{incorrect}} \gets \mathcal{N}(\|e_i\|; \mu^{\text{incorrect}}_i, \sigma^{\text{incorrect}}_i)$ \Comment{\textcolor{cyan}{Prob. under incorrect dist.}}
    
    \If{$p_{\text{correct}} > p_{\text{incorrect}}$} \Comment{\textcolor{cyan}{Global norm condition}}
        \State Compute softmax logits $l_i$ at exit $EE_i$
        \State Predict class $\hat{c} \gets \arg\max(l_i)$
        
        \If{class $\hat{c}$ has class-specific statistics}
            \State $p_{\text{correct},\hat{c}} \gets \mathcal{N}(\|e_i\|; \mu^{\text{correct}}_{i,\hat{c}}, \sigma^{\text{correct}}_{i,\hat{c}})$
            \State $p_{\text{incorrect},\hat{c}} \gets \mathcal{N}(\|e_i\|; \mu^{\text{incorrect}}_{i,\hat{c}}, \sigma^{\text{incorrect}}_{i,\hat{c}})$
            
            \If{$p_{\text{correct},\hat{c}} > p_{\text{incorrect},\hat{c}}$} \Comment{\textcolor{cyan}{Class-specific norm condition}}
                \State \Return prediction $\hat{c}$ from exit $EE_i$
            \EndIf
        \Else
            \State \Return prediction $\hat{c}$ from exit $EE_i$ 
        \EndIf
    \EndIf
\EndFor

\State \Return prediction from exit  $EE_{N-1}$ \Comment{\textcolor{cyan}{Final exit if no EE triggered}}
\end{algorithmic}
}
\end{algorithm}
The core of our triggering criteria is to model the distributions of embedding norms for correct and incorrect predictions. Using a reference set, we pre-calculate the mean and standard deviation of the spatial norms $\|\tilde{p}\|$ for both correct and incorrect predictions at each exit gate. This calibration is done both globally across all classes and on a per-class basis. During inference, a sample is processed sequentially through the exits. At each gate, its embedding norm is evaluated using a two-stage probabilistic check: 
(1) \textbf{Global norm condition:} We compute the probability of the sample's norm under the pre-computed global Gaussian distributions for correct and incorrect predictions. If the norm is more probable under the `correct' distribution, the sample becomes a candidate for an early exit;
(2) \textbf{Class-specific norm condition:} If the global check passes, we make a preliminary class prediction. We then perform a second, more stringent check using the norm distributions specific to that predicted class. If this check also passes, the model confidently exits with the prediction.
If a sample fails either check, it is deemed uncertain and is passed to the next, more powerful exit stage. This creates a highly selective trigger that only allows high-confidence samples to exit early.

We simulate this triggering strategy on the ESC-50 validation set using our trained \modelname{} with MobileNetV3 backbone. The results, summarized in Table \ref{tab:early_exit_summary}, show our \textbf{Class-specific norm Exit} strategy achieves an overall accuracy of \textbf{87.75\%}, significantly outperforming not only a standard entropy-based trigger (70.83\%) but also the powerful \textbf{Final Exit Only} (non-EE) baseline (83.39\%). This is achieved while saving \textbf{36.1\%} of the Multiply-Accumulate operations (MACs), relative to the Final Exit Only (non-EE) baseline, demonstrating an important outcome where a model becomes both more accurate and more efficient.

\begin{table}[t]
\centering
\vspace{-1em}
\caption{Early-Exit Triggering Results for Audio Tagging task with two early-exit gates. Samples shown as \% of total queries.}
\label{tab:early_exit_summary}
\resizebox{\columnwidth}{!}{%
\begin{tabular}{c|ccc|c|c}
\toprule
\textbf{Exit Strategy} & \textbf{${EE}_{0}$ \%} & \textbf{${EE}_{1}$ \%} & \textbf{Final \%} & \textbf{MACs saved \%} & \textbf{Accuracy \%} \\
\midrule
\rowcolor{gray!20}Final Exit Only      & --   & --   & 100.0   & --   & 83.39 \\
\rowcolor{gray!20}Exit at ${EE}_{1}$          & --   & 100.0 & --     & 44.3 & 71.32 \\
\rowcolor{gray!20}Exit at ${EE}_{0}$          & 100.0 & --   & --     & 62.5 & 62.08 \\
\midrule
Entropy (\baseline{})     & 47.19  & 12.67  & 40.14    & 35.1 & 70.83 \\
\midrule
Global norm Exit (\modelname{})    & 35.6  & 36.7  & 27.6    & \cellcolor{green!20}{\textbf{38.5}} & \cellcolor{green!20}{74.02} \\
Class-specific norm Exit (\modelname{}) & 30.1  & 39.1  & 30.9    & \cellcolor{green!20}{36.1} & \cellcolor{green!20}{\textbf{87.75}} \\
\bottomrule
\end{tabular}
}
\vspace{-1em}
\end{table}

%% file: sec/5_conclusion.tex
\section{Conclusion}
In this work, we addressed the critical challenge of designing efficient and reliable multi-stage event detection systems by introducing \modelname{}, a novel framework that leverages hyperbolic geometry to model the hierarchical structure within Early-Exit networks. 
By employing a novel entailment loss across the exits, \modelname{} learns a joint latent space where the geometric distance from the origin serves as a principled and robust measure of model uncertainty, ensuring that deeper network layers systematically refine the representations of shallower ones.
Our experiments
demonstrate that \modelname{} significantly outperforms standard Euclidean Early-Exit baselines, particularly at the earliest, low-compute exits. 
We showed that the resulting hyperbolic space is more parameter-efficient and enables a novel, geometry-aware triggering mechanism that achieves a superior accuracy-efficiency trade-off, even surpassing the performance of a final-exit-only model. 
We validate that treating uncertainty as a geometric property is a powerful paradigm for Early-Exit neural networks, opening promising avenues for developing more robust and context-aware intelligent systems for real-world, resource-constrained applications.

%% file: sec/supple_motivation_delta.tex
\section{Motivation for Hyperbolic Shift in Early-Exits}
\label{sec:appendix_motivation_delta}

The premise of our work is that the representations learned by deep backbone networks are inherently hierarchical across their depth. We analyze the geometric structure of intermediate embeddings from a pre-trained BEATs \cite{chen2022beats} audio backbone. We adopt the concept of 
Gromov's $\delta$-hyperbolicity \cite{gromov1987hyperbolic}, a formal measure that quantifies the \textit{tree-likeness} of a metric space. A low, scale-invariant $\delta$-hyperbolicity value, denoted 
$\delta_{rel} \in [0,1]$ \footnote{$\delta_{rel} = \frac{2\delta}{\text{diameter}}$ \\
Diameter: maximal pairwise distance. 
Any latent space is considered $\delta$-hyperbolic if, for some value $\delta$, every point located on the edge of a geodesic triangle is within a distance of $\delta$ from another edge.},
indicates that the space is highly tree-like and thus well-suited for embedding in a hyperbolic geometry \cite{khrulkov2020hyperbolic}.

We conduct an experiment where we extract embeddings from the backbone at different depths: 25\% through the network, 50\% through, and at the final layer (100\%). We then compute $\delta_{rel}$
  both within the set of embeddings from a single layer (\ie{} intra-layer) and between the sets of embeddings from different layers (\ie{} inter-layer). The results, summarized in Table \ref{tab:hyperbolicity_measures}, reveal two key findings. First, the intra-layer embeddings at each depth exhibit low $\delta_{rel}$ 
  values (0.23-0.30), confirming that the representations for different audio samples are already organized in a hierarchical order. More importantly, the inter-layer hyperbolicity is even more pronounced, with $\delta_{rel}$
  values as low as 0.143 between the 50\% and 100\% layers.

This strong empirical evidence suggests that a natural hierarchical structure exists not just among audio samples (inline with \cite{khrulkov2020hyperbolic}'s observation for image samples), but critically, across the depth of the audio backbone itself. The representations at deeper layers are structurally related to those at shallower layers in a tree-like manner. This finding motivates our core proposal: 
to replace the geometrically unstructured (hierarchy) Euclidean space of traditional Early-Exit models with a hyperbolic latent space, 
which provides a natural inductive bias for learning and preserving these hierarchical relationships.

\begin{table}[htbp]
\centering
\caption{Gromov's $\delta$-hyperbolicity for intermediate embeddings from a pre-trained BEATs backbone. We compare both intra-layer (top) and inter-layer (bottom) configurations. The significantly lower $\delta_{rel}$ values for inter-layer comparisons indicate a strong hierarchical structure across the network's depth and strongly motivate the use of hyperbolic geometry to model the network's depth-wise progression.}
\label{tab:hyperbolicity_measures}
\begin{tabular}{c|c|c|c}
\toprule
\textbf{X} & \textbf{Y} & $\delta_{rel}$ & $c$ \\
\midrule
25\%  & 25\%  & 0.282 & 0.26  \\
50\%  & 50\%  & 0.304 & 0.223 \\
100\% & 100\% & 0.233 & 0.379 \\
\midrule
25\%  & 50\%  & 0.247 & 0.338 \\
25\%  & 100\% & 0.148 & 0.94  \\
50\%  & 100\% & 0.143 & 1.012 \\
\bottomrule
\end{tabular}

\end{table}

%% file: sec/supple_early_exit_analysis.tex
\section{Detailed Early-Exit Trigger Analysis}
\label{sec:appendix_ee_trigger_detailed}

We further detail our breakdown of proposed EE triggers (Section \ref{sec:trigger}) in Table \ref{tab:early_exit_detailed_analysis}. It is evident that a
geometric trigger is exceptionally precise at identifying samples it can classify correctly: 
of the samples exited at ${EE}_{0}$ and ${EE}_{1}$, over 98.8\% and 99.7\% are classified correctly, respectively. 
The model intelligently offloads the truly difficult samples (approx. 31\% of the total) to the final, most capable exit. 
This demonstrates that our geometry-aware triggering mechanism successfully operationalizes the learned hierarchy, completing the \modelname{} framework and delivering a superior accuracy-efficiency trade-off.

\begin{table}[t]
\centering
\caption{Detailed Early-Exit Analysis for Global Norm Exit and Classwise Norm Exit Strategies}
\label{tab:early_exit_detailed_analysis}
\resizebox{\columnwidth}{!}{%
\begin{tabular}{c|c|c|c|c}
\toprule
\textbf{Exit Strategy} & \textbf{Gate} & \textbf{Triggered \%} & \textbf{Correct \%} & \textbf{Incorrect \%} \\
\midrule
\multirow{3}{*}{Global Norm Exit} 
& ${EE}_{0}$  &  35.61 & 80.52 & 19.48 \\
& ${EE}_{1}$  & 36.74 & 78.13 & 21.87 \\
& Final & 27.65 & 60.18 & 39.82 \\
\midrule
\multirow{3}{*}{Classwise Norm Exit} 
& ${EE}_{0}$  & 30.05 & 98.82 & 1.18 \\
& ${EE}_{1}$  & 39.08 & 99.73 & 0.27 \\
& Final & 30.87& 61.81 & 38.19 \\
\bottomrule
\end{tabular}
}
\end{table}

%% file: sec/supple_additional_vis.tex
\section{Additional Qualitative visualizations}
\label{sec:appendix_viz}

\subsection{UMAP Visualization of Exit Gate Embeddings}
\begin{figure}[t]
    \centering
    \resizebox{\columnwidth}{!}{%
        \includegraphics{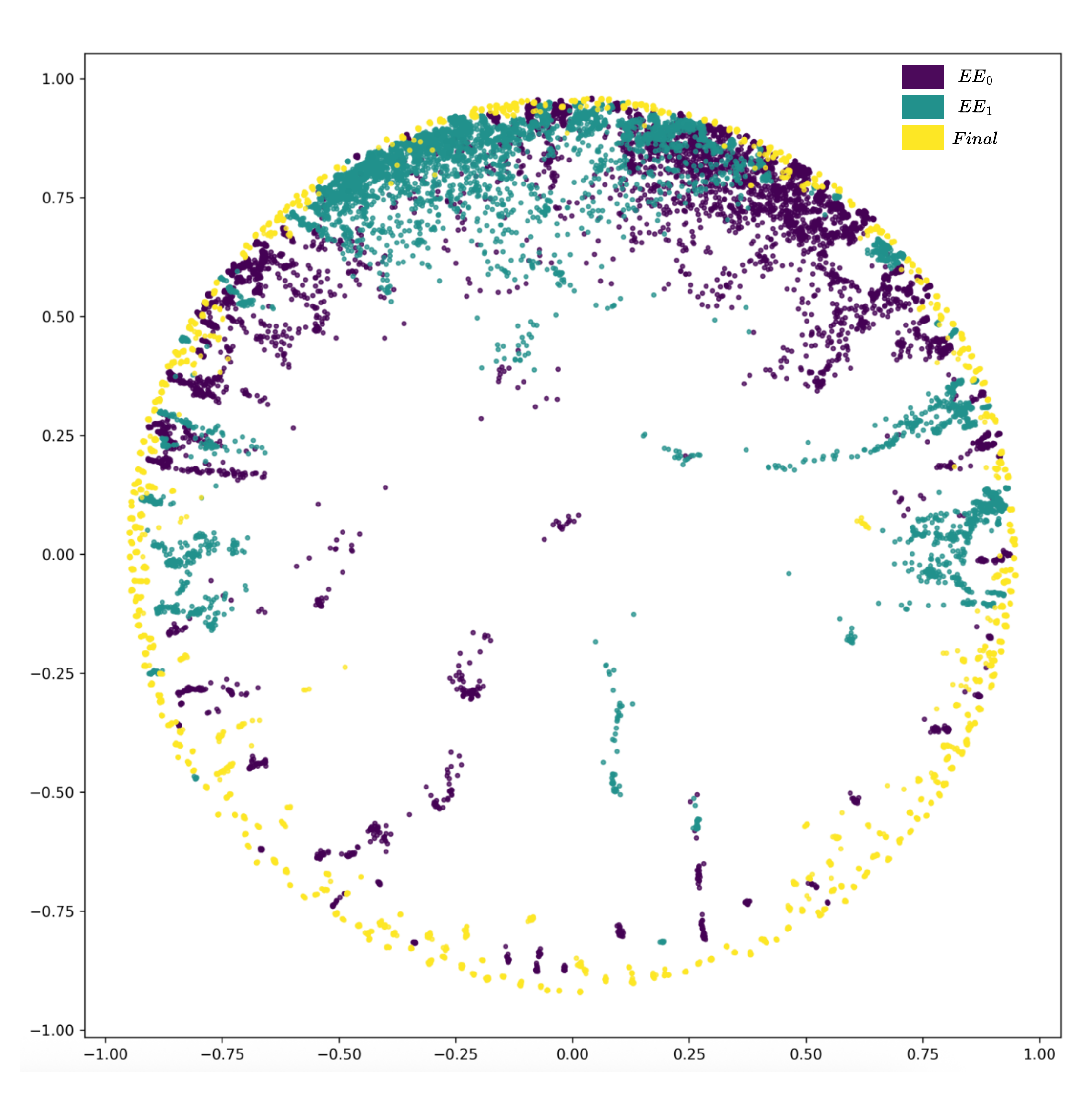}
    }

        \caption{UMAP \cite{mcinnes2018umap} visualization of the learned hyperbolic embeddings from the SED model, projected onto the \emph{Poincaré} disk. The embeddings are colored by their exit level (${EE}_{0}$, ${EE}_{1}$, $Final$). The plot shows a clear radial hierarchy, with earlier exit embeddings positioned more centrally, providing evidence of the learned entailment structure.}

    \label{fig:ee_umap}
\end{figure}
In addition to the t-SNE plots in Section \ref{fig:qualitative_main_paper}, we use UMAP (Uniform Manifold Approximation and Projection) \cite{mcinnes2018umap} to visualize the learned embeddings, as shown in Fig. \ref{fig:ee_umap}. 
The embeddings from the three exit gates are projected from the \emph{Lorentz} hyperboloid onto its equivalent \emph{Poincaré} disk representation. The visualization, colored by exit level, provides further evidence of the hierarchical structure imposed by our entailment loss. The embeddings from the first exit, ${EE}_{0}$ (\textcolor{purple}{purple}), are predominantly located in the central region of the disk, representing higher uncertainty. The embeddings from the second exit, ${EE}_{1}$ (\textcolor{teal}{teal}), extend outwards from this core, and the $Final$ exit embeddings (\textcolor{yellow}{yellow}) are pushed furthest towards the periphery. This clear radial separation confirms that the model learns a structured progression from general to specific representations across its depth.

\subsection{Contextual Clustering with Hyperbolic k-means}

\begin{figure*}[t]
    \centering
    \includegraphics[width=1.9\columnwidth]{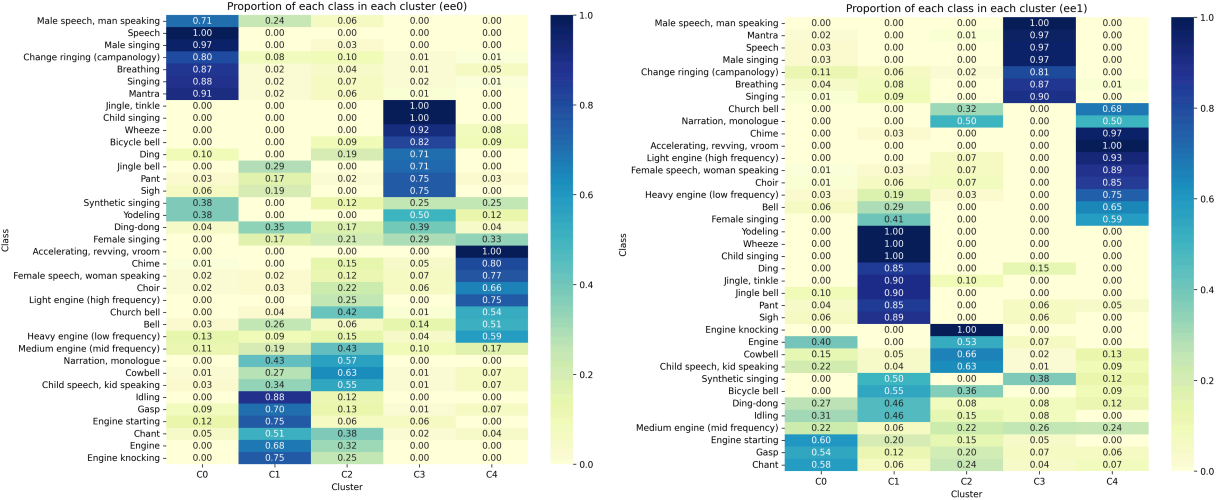}
        \caption{Proportion of hand-picked Audioset Strong classes within each of the 5 clusters discovered by hyperbolic k-means, for embeddings from ${EE}_{0}$ (left) and ${EE}_{1}$ (right). The unsupervised clusters show a strong correspondence with high-level acoustic concepts (e.g., human speech, bells, engines), indicating that the Early-Exits learn a meaningful contextual hierarchy.}

    \label{fig:hyp_kmeans}
\end{figure*}

To investigate the semantic organization of the learned space at the Early-Exits, we perform an unsupervised clustering experiment.
Our hypothesis is that the Early-Exits learn to group sounds into broader, contextually relevant acoustic categories, even without explicit supervision to do so.
Specifically, we select five distinct, high-level acoustic concepts from the Audioset Strong evaluation set: Respiratory, Ringing, Speech, Singing, and Mechanical (engines), each comprising several fine-grained classes. We gather evaluation samples belonging to these classes and apply hyperbolic k-means clustering (k=5) to their embeddings taken from ${EE}_{0}$ and ${EE}_{1}$ exits, separately.
Fig. \ref{fig:hyp_kmeans} shows the proportion of each fine-grained class within the emergent clusters found by k-means. The results reveal a remarkable correspondence between the unsupervised clusters and our predefined semantic groups. For example, at ${EE}_{0}$, Cluster $C0$ is overwhelmingly composed of various speech and singing classes (human vocalizations), while Cluster $C3$ is almost exclusively made up of different types of bell and chime sounds (high-frequency alerts/musical sounds). Similarly, a significant portion of engine-related sounds is grouped into Cluster $C2$.

\noindent\textbf{Implication for Contextual AI.} This emergent clustering demonstrates that the Early-Exits in \modelname{} learn a meaningful acoustic taxonomy. ${EE}_{0}$ can effectively distinguish between high-level concepts like ``human vocalizations'' or ``mechanical noise'' even if it remains uncertain about the specific subclass. This capability is highly valuable for contextual AI on resource-constrained devices. An ``always-on'' system could use a computationally cheap Early-Exit to make a broad contextual inference (e.g., ``human presence detected,'' ``vehicle nearby'') and only trigger the more expensive, deeper layers when a fine-grained classification is required, enabling a more intelligent and efficient allocation of resources \cite{smith2023reshaping}.

%% file: sec/supple_lookahead.tex
\section{Lookaheads within Entailment Cones}
\label{sec:appendix_lookahead}

\begin{figure}[htbp]
    \centering
    \includegraphics[width=0.8\columnwidth]{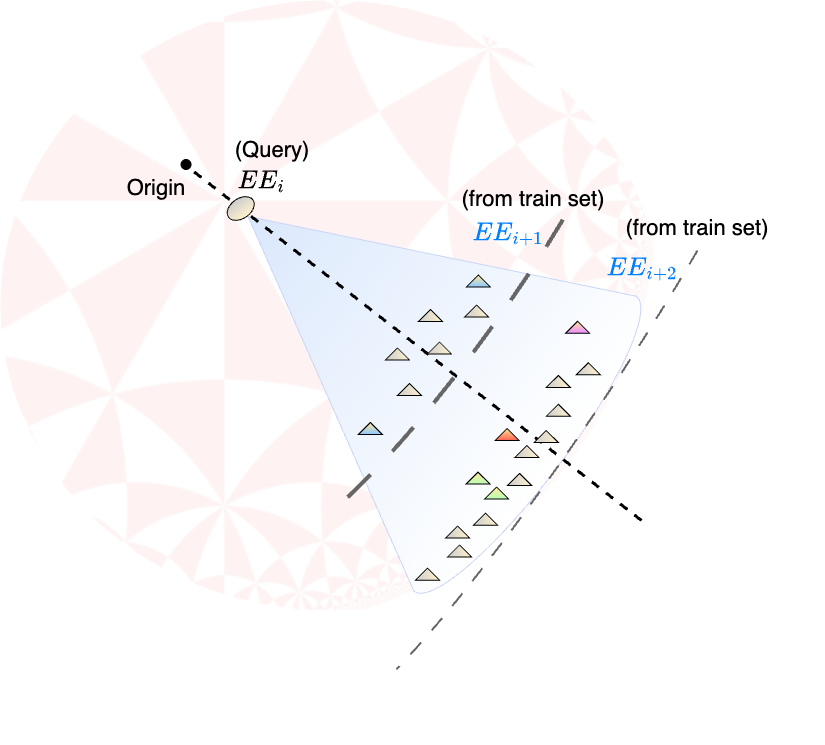}
    \caption{A conceptual illustration of the \textit{look ahead} prediction strategy. A \texttt{query} sample's embedding at an Early-Exit, ${EE}_{i}$, defines an entailment cone. We \textit{look ahead} by identifying \texttt{reference} embeddings from the training set of subsequent exits (e.g., ${EE}_{i+1}$, $Final$) that fall within this cone. The ground-truth classes of these retrieved reference samples are then used to forecast the \texttt{query}'s most likely class.}
    \label{fig:lookahead_concept}
\end{figure}

In Algorithm 1, we demonstrated a triggering mechanism based on the norm of hyperbolic embeddings, which serves as a proxy for uncertainty. Beyond this, we explore whether the entailment cone itself—the core of our hierarchical training objective—could be directly harnessed for tirggering inference. Inspired by work on predicting uncertain futures \cite{suris2021learning}, where hyperbolic models \textit{hedge their bets} by forecasting a more abstract outcome, we investigate if an embedding at an Early-Exit, $h_i$, could \textit{forecast} its final classification by examining the classes of more refined embeddings that are geometrically consistent with it (\ie{} fall in its entailment cone).

We design an experiment where each sample from the ESC-50 validation set acts as a \texttt{query} represented by its embedding at the first exit, ${EE}_{0}$. A \texttt{reference} set consists of all training set embeddings from the subsequent, more refined exits (${EE}_{1}$ and $Final$). For each \texttt{query}, we identify all \texttt{reference} embeddings that fall within its entailment cone, a process conceptually illustrated in Fig. \ref{fig:lookahead_concept}. Since the entailment loss is non-zero during our training, we relax the strict condition with a threshold $T$, such that a \texttt{reference} sample $h_{ref}$ is considered to be within the cone of a query $h_{query}$ if $ext(h_{query}, h_{ref}) \le T \cdot aper(h_{query})$.

\begin{figure}[htbp]
    \centering
    \includegraphics[width=\columnwidth]{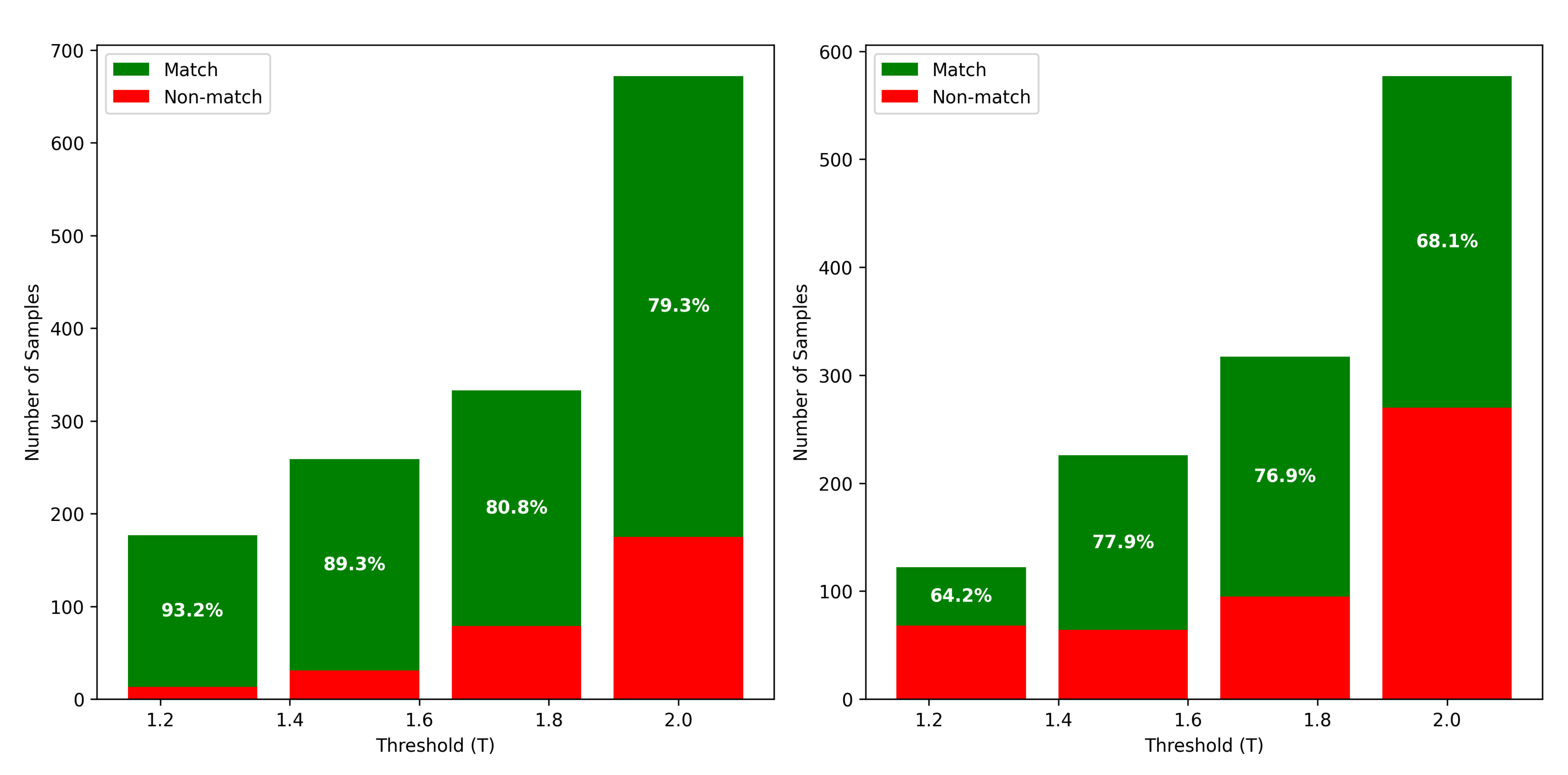}
    \caption{Results of the \textit{look ahead} prediction experiment. For different entailment cone thresholds ($T$), we show the number of retrieved \texttt{reference} samples from later exits that either match (green) or do not match (red) the \texttt{query} sample's ground-truth class. The percentages indicate the precision (match / total retrieved). The left and right plots correspond to using \texttt{reference} samples from ${EE}_{1}$ and the $Final$ exit, respectively.}
    \label{fig:lookahead_vis}
\end{figure}

Fig. \ref{fig:lookahead_vis} shows that at tight thresholds (e.g., $T=1.2$), the precision is remarkably high: 93.2\% of the \texttt{reference} samples retrieved from ${EE}_{1}$ share the same ground-truth class as the \texttt{query} sample. This indicates that the entailment cone is semantically coherent and contains strong predictive information about the \texttt{query}'s identity. As the threshold is relaxed, the number of retrieved samples increases, but precision naturally decreases.
While promising, we present this as an exploratory analysis rather than a practical inference algorithm due to two main challenges. First, the computational cost of comparing a \texttt{query} against a large reference set is prohibitive for real-time applications. Second, some \texttt{query} samples do not retrieve any \texttt{reference} samples at stricter thresholds, limiting the coverage of the method.
However, this exploration successfully validates the rich, predictive structure of the \modelname{} latent space and opens several exciting avenues for future work. A key direction would be to develop methods to make this \textit{look ahead} approach practical, perhaps by learning a small, representative set of prototype \texttt{reference} embeddings to reduce the search space, or by training a model to directly predict the class distribution within an embedding's entailment cone. Our initial result strongly suggests that the geometry learned by \modelname{} is not just a representational artifact, but a potentially powerful tool for future inference strategies.

%% file: sec/supple_traversal.tex
\section{Traversing Along the Learned Hierarchy}
\label{sec:appendix_traversal}

To evaluate the hierarchical structure learned by \modelname{}, we conduct a traversal experiment inspired by recent work in hyperbolic representation learning \cite{desai2023hyperbolic, pal2024compositional} with an objective to analyze the path from a specific, fine-grained embedding (from the Final exit) to the most general concept in the hyperbolic latent space (the \texttt{[ROOT]}). A well-structured hierarchy should reveal a smooth progression from specific to abstract concepts along this path.

\begin{table*}[t]
\centering
\caption{Traversal paths from a query audio embedding to the \texttt{[ROOT]}. The path for \modelname{} shows a clear hierarchical progression from specific to general concepts, while the \baseline{} path collapses almost immediately. The colors indicate the exit-gate that the retrieved embedding belongs to.}
\label{tab:traversal_results}
\resizebox{\textwidth}{!}{%
\begin{tabular}{@{}lll@{}}
\toprule
\textbf{Query' Class (Exit)} & \textbf{\modelname{} Traversal Path} & \textbf{\baseline{} Traversal Path} \\ \midrule
Clapping (\texttt{Final})  & 
    \texttt{\textcolor{yellow}{Clapping}({Final})} $\rightarrow$ 
    \texttt{\textcolor{teal}{Clapping}($EE_1$)} $\rightarrow$ 
    \texttt{\textcolor{purple}{Clapping}($EE_0$)} $\rightarrow$ \dots $\rightarrow$ 
    \texttt{[ROOT]} & 
    \texttt{\textcolor{purple}{Clapping}($EE_0$)} $\rightarrow$ 
    \texttt{[ROOT]} \\
\addlinespace[0.5em]
Church Bells (\texttt{Final})& 
    \texttt{\textcolor{yellow}{Church Bells}({Final})} $\rightarrow$ 
    \texttt{\textcolor{teal}{Church Bells}($EE_1$)} $\rightarrow$ 
    \texttt{\textcolor{purple}{Cow}($EE_0$)} $\rightarrow$ \dots $\rightarrow$ 
    \texttt{[ROOT]} & 
    \texttt{\textcolor{teal}{Church Bells}($EE_1$)} $\rightarrow$ 
    \texttt{[ROOT]} \\
\addlinespace[0.5em]
Wood Hammer (\texttt{Final})& 
    \texttt{\textcolor{teal}{Wood Hammer}($EE_1$)} $\rightarrow$ \dots $\rightarrow$ 
    \texttt{\textcolor{purple}{Glass Breaking}($EE_0$)} $\rightarrow$ \dots $\rightarrow$ 
    \texttt{[ROOT]} & 
    \texttt{\textcolor{yellow}{Wood Hammer}({Final})} $\rightarrow$ 
    \texttt{[ROOT]} \\
\addlinespace[0.5em]
Vacuum Cleaner (\texttt{Final})&
    \texttt{\textcolor{teal}{Vacuum Cleaner}($EE_1$)} $\rightarrow$ 
    \texttt{\textcolor{purple}{Vacuum Cleaner}($EE_0$)} $\rightarrow$ \dots $\rightarrow$ 
    \texttt{[ROOT]} &
    \texttt{\textcolor{yellow}{Vacuum Cleaner}({Final})} $\rightarrow$ 
    \texttt{[ROOT]} \\
\bottomrule
\end{tabular}%
}
\end{table*}


\textbf{\texttt{[ROOT]} Embedding.} For \modelname{}, the \texttt{[ROOT]} of the hierarchy is naturally defined as the origin of the Lorentz hyperboloid, $o \in \mathbb{L}_{c}^{n}$, as it entails the entire representation space. For the Euclidean baseline (\baseline{}), which lacks a natural origin on its hypersphere, we empirically define the \texttt{[ROOT]} as the centroid of all training data embeddings, which is then $L_2$-normalized.

\textbf{Traversal via Interpolation.} We traverse the latent space by interpolating 50 steps along the shortest path (geodesic) between a query audio's final-exit embedding, $h_{Final}$, and the \texttt{[ROOT]}.
\begin{itemize}
    \item \textbf{For \modelname{}}, this is achieved by first mapping $h_{Final}$ to the tangent space at the origin via the logarithmic map ($v = \text{logm}_{o}(h_{Final})$). We then perform linear interpolation in this flat tangent space between $v$ and the origin. Each interpolated vector is subsequently mapped back onto the hyperboloid using the exponential map ($h_{interp} = \text{expm}_{o}(v_{interp})$).
    \item \textbf{For \baseline{}}, we perform standard linear interpolation (LERP) \cite{desai2023hyperbolic} between the $L_2$-normalized embeddings of $h_{Final}$ and the \texttt{[ROOT]}, followed by re-normalization at each step.
\end{itemize}

\textbf{Nearest Neighbor Retrieval.} At each of the 50 interpolated steps, we perform a nearest-neighbor search. The reference set for this search consists of all embeddings from the training dataset across all three exit levels ($EE_0$, $EE_1$, and $EE_{Final}$). For \modelname{}, similarity is measured by the Lorentzian inner product, while for \baseline{}, it is measured by cosine similarity.

The results of the traversal experiment, summarized in Table~\ref{tab:traversal_results}, reveal a stark contrast between the latent spaces learned by \modelname{} and the Euclidean baseline.
\modelname{} consistently reveals a structured, multi-step traversal path that reflects the intended model hierarchy. The path progresses logically from neighbors in the specific, high-resolution $EE_{Final}$ or $EE_1$ space to neighbors in the more general, low-resolution $EE_0$ space before converging at the \texttt{[ROOT]}. This provides strong qualitative evidence that our entailment loss successfully organizes the embeddings according to their stage of refinement.

Most notably, the traversal reveals that the earliest exit, $EE_0$, learns an emergent acoustic taxonomy. For instance, when traversing from a ``Wood Hammer'' query (a sharp, percussive sound), the nearest neighbor at the $EE_0$ level is ``Glass Breaking,'' a semantically distinct but acoustically similar transient event. 
The earliest exit learns to group sounds by their broader acoustic morphology, a more general concept than their specific semantic label, which is precisely the desired behavior of a hierarchical system.
In contrast, the traversals within the Euclidean latent space lack this rich structure. In all tested cases, the path collapses to the \texttt{[ROOT]} after retrieving at most one neighbor. This suggests the \baseline{} space is not organized in a navigable, nested hierarchy, further underscoring the benefits of the geometric inductive bias provided by hyperbolic space for training Early-Exit networks.

Beyond providing qualitative validation of the learned hierarchy, these findings point towards several practical applications for the structured latent space learned by \modelname{}. The navigable hierarchy offers a powerful tool for model interpretability and error analysis, allowing to trace the refinement process for a given input. Furthermore, the emergent acoustic taxonomy at the earliest exit could enable more sophisticated, context-aware triggering mechanisms. For instance, an ``always-on'' device could use the computationally cheap $EE_0$ to make broad contextual inferences (e.g., detecting a ``transient event'') and only activate the deeper, more power-intensive exits when a fine-grained classification is necessary. This opens avenues for designing more efficient and intelligent sensing systems that leverage a deeper understanding of their acoustic environment.

%% file: Template.bbl
\begin{thebibliography}{10}

\bibitem{dorschky2015framework}
Eva Dorschky, Dominik Schuldhaus, Harald Koerger, and Bjoern~M Eskofier,
\newblock ``A framework for early event detection for wearable systems,''
\newblock in {\em ACM International Symposium on Wearable Computers}, 2015.

\bibitem{shankar2024edge}
Vasuki Shankar,
\newblock ``Edge ai: a comprehensive survey of technologies, applications, and challenges,''
\newblock in {\em IEEE ACET}, 2024.

\bibitem{Prince2019DeployingAcoustic}
Peter Prince, Andrew Hill, Evelyn Pi{\~n}a~Covarrubias, Patrick Doncaster, Jake~L Snaddon, and Alex Rogers,
\newblock ``Deploying acoustic detection algorithms on low-cost, open-source acoustic sensors for environmental monitoring,''
\newblock {\em Sensors}, 2019.

\bibitem{panda2016conditional}
Priyadarshini Panda, Abhronil Sengupta, and Kaushik Roy,
\newblock ``Conditional deep learning for energy-efficient and enhanced pattern recognition,''
\newblock in {\em IEEE DATE}, 2016.

\bibitem{Zhang2021BasisNet}
Mingda Zhang, Chun-Te Chu, Andrey Zhmoginov, Andrew Howard, Brendan Jou, Yukun Zhu, Li~Zhang, Rebecca Hwa, and Adriana Kovashka,
\newblock ``Basisnet: Two-stage model synthesis for efficient inference,''
\newblock in {\em IEEE CVPR}, 2021.

\bibitem{rahmath2024early}
Haseena Rahmath~P, Vishal Srivastava, Kuldeep Chaurasia, Roberto~G Pacheco, and Rodrigo~S Couto,
\newblock ``Early-exit deep neural network-a comprehensive survey,''
\newblock {\em ACM Computing Surveys}, 2024.

\bibitem{Vendrov2015OrderEmbeddings}
Ivan Vendrov, Ryan Kiros, Sanja Fidler, and Raquel Urtasun,
\newblock ``Order-embeddings of images and language,''
\newblock {\em ICLR}, 2016.

\bibitem{nguyen2015deep}
Anh Nguyen, Jason Yosinski, and Jeff Clune,
\newblock ``Deep neural networks are easily fooled: High confidence predictions for unrecognizable images,''
\newblock in {\em IEEE CVPR}, 2015.

\bibitem{guo2017calibration}
Chuan Guo, Geoff Pleiss, Yu~Sun, and Kilian~Q Weinberger,
\newblock ``On calibration of modern neural networks,''
\newblock in {\em ICML}, 2017.

\bibitem{nickel2017poincare}
Maximillian Nickel and Douwe Kiela,
\newblock ``Poincar{\'e} embeddings for learning hierarchical representations,''
\newblock {\em NeurIPS}, 2017.

\bibitem{mathieu2019continuous}
Emile Mathieu, Charline Le~Lan, Chris~J Maddison, Ryota Tomioka, and Yee~Whye Teh,
\newblock ``Continuous hierarchical representations with poincar{\'e} variational auto-encoders,''
\newblock {\em NeurIPS}, 2019.

\bibitem{ganeaentailmentcone}
Octavian Ganea, Bécigneul Gary, and Hofmann Thomas,
\newblock ``Hyperbolic entailment cones for learning hierarchical embeddings,''
\newblock in {\em ICML}, 2018.

\bibitem{mettes2024hyperbolic}
Pascal Mettes, Mina Ghadimi~Atigh, Martin Keller-Ressel, Jeffrey Gu, and Serena Yeung,
\newblock ``Hyperbolic deep learning in computer vision: A survey,''
\newblock {\em IJCV}, 2024.

\bibitem{desai2023hyperbolic}
Karan Desai, Maximilian Nickel, Tanmay Rajpurohit, Justin Johnson, and Shanmukha~Ramakrishna Vedantam,
\newblock ``Hyperbolic image-text representations,''
\newblock in {\em ICML}, 2023.

\bibitem{pal2024compositional}
Avik Pal, Max van Spengler, Guido Maria~D'Amely di~Melendugno, Alessandro Flaborea, Fabio Galasso, and Pascal Mettes,
\newblock ``Compositional entailment learning for hyperbolic vision-language models,''
\newblock {\em ICLR}, 2025.

\bibitem{sala2018representation}
Frederic Sala, Chris De~Sa, Albert Gu, and Christopher R{\'e},
\newblock ``Representation tradeoffs for hyperbolic embeddings,''
\newblock in {\em ICML}, 2018.

\bibitem{tifrea2018poincar}
Alexandru Tifrea, Gary B{\'e}cigneul, and Octavian-Eugen Ganea,
\newblock ``Poincar$\backslash$'e glove: Hyperbolic word embeddings,''
\newblock {\em arXiv preprint arXiv:1810.06546}, 2018.

\bibitem{peng2021hyperbolic}
Wei Peng, Tuomas Varanka, Abdelrahman Mostafa, Henglin Shi, and Guoying Zhao,
\newblock ``Hyperbolic deep neural networks: A survey,''
\newblock {\em IEEE TPAMI}, 2021.

\bibitem{bdeir2023fully}
Ahmad Bdeir, Kristian Schwethelm, and Niels Landwehr,
\newblock ``Fully hyperbolic convolutional neural networks for computer vision,''
\newblock {\em ICLR}, 2024.

\bibitem{he2025hyperbolic}
Neil He, Hiren Madhu, Ngoc Bui, Menglin Yang, and Rex Ying,
\newblock ``Hyperbolic deep learning for foundation models: A survey,''
\newblock in {\em ACM SIGKDD}, 2025.

\bibitem{liu2019hyperbolic}
Qi~Liu, Maximilian Nickel, and Douwe Kiela,
\newblock ``Hyperbolic graph neural networks,''
\newblock {\em NeurIPS}, 2019.

\bibitem{petermann2023hyperbolic}
Darius Petermann, Gordon Wichern, Aswin Subramanian, and Jonathan Le~Roux,
\newblock ``Hyperbolic audio source separation,''
\newblock in {\em IEEE ICASSP}, 2023.

\bibitem{petermann2024hyperbolic}
Darius Petermann and Minje Kim,
\newblock ``Hyperbolic distance-based speech separation,''
\newblock in {\em IEEE ICASSP}, 2024.

\bibitem{germain2023hyperbolic}
Fran{\c{c}}ois~G Germain, Gordon Wichern, and Jonathan Le~Roux,
\newblock ``Hyperbolic unsupervised anomalous sound detection,''
\newblock in {\em IEEE WASPAA}. IEEE, 2023, pp. 1--5.

\bibitem{hong2023hyperbolic}
Jie Hong, Zeeshan Hayder, Junlin Han, Pengfei Fang, Mehrtash Harandi, and Lars Petersson,
\newblock ``Hyperbolic audio-visual zero-shot learning,''
\newblock in {\em IEEE ICCV}, 2023.

\bibitem{cramer2020chirping}
Aurora~Linh Cramer, Vincent Lostanlen, Andrew Farnsworth, Justin Salamon, and Juan~Pablo Bello,
\newblock ``Chirping up the right tree: Incorporating biological taxonomies into deep bioacoustic classifiers,''
\newblock in {\em IEEE ICASSP}, 2020.

\bibitem{khandelwal2023multi}
Tanmay Khandelwal and Rohan~Kumar Das,
\newblock ``A multi-task learning framework for sound event detection using high-level acoustic characteristics of sounds,''
\newblock {\em Interspeech}, 2023.

\bibitem{liang2024learning}
Jinhua Liang, Huy Phan, and Emmanouil Benetos,
\newblock ``Learning from taxonomy: Multi-label few-shot classification for everyday sound recognition,''
\newblock in {\em IEEE ICASSP}, 2024.

\bibitem{Mishne2023NumericalHyperbolic}
Gal Mishne, Zhengchao Wan, Yusu Wang, and Sheng Yang,
\newblock ``The numerical stability of hyperbolic representation learning,''
\newblock in {\em ICML}, 2023.

\bibitem{Minkowski1908RaumZeit}
Hermann Minkowski,
\newblock ``Raum und zeit,''
\newblock {\em Physikalische Zeitschrift}, 1908.

\bibitem{Einstein2015PrincipleRelativity}
Albert Einstein, H.~A. Lorentz, Hermann Minkowski, and Hermann Weyl,
\newblock {\em The Principle of Relativity: A Collection of Original Memoirs on the Special and General Theory of Relativity},
\newblock Martino Fine Books, 2nd edition, 2015.

\bibitem{Law2019Lorentzian}
Marc~T Law, Renjie Liao, Jake Snell, and Richard~S Zemel,
\newblock ``Lorentzian distance learning for hyperbolic representations,''
\newblock in {\em ICML}, 2019, pp. 3672--3681.

\bibitem{Nickel2018LorentzHyperbolic}
Maximilian Nickel and Douwe Kiela,
\newblock ``Learning continuous hierarchies in the lorentz model of hyperbolic geometry,''
\newblock in {\em ICML}, 2018.

\bibitem{chen2022beats}
Sanyuan Chen, Yu~Wu, Chengyi Wang, Shujie Liu, Daniel Tompkins, Zhuo Chen, and Furu Wei,
\newblock ``{BEATs}: Audio pre-training with acoustic tokenizers,''
\newblock {\em ICML}, 2022.

\bibitem{koonce2021mobilenetv3}
Brett Koonce,
\newblock ``Mobilenetv3,''
\newblock in {\em Convolutional neural networks with swift for tensorflow: image recognition and dataset categorization}. Springer, 2021.

\bibitem{kubatytrain}
Piotr Kubaty, Bartosz W{\'o}jcik, Bart{\l}omiej~Tomasz Krzepkowski, Monika Michaluk, Tomasz Trzcinski, Jary Pomponi, and Kamil Adamczewski,
\newblock ``How to train your multi-exit model? analyzing the impact of training strategies,''
\newblock in {\em ICML}, 2025.

\bibitem{piczak2015esc}
Karol~J Piczak,
\newblock ``{ESC}: Dataset for environmental sound classification,''
\newblock in {\em ACM Multimedia}, 2015.

\bibitem{Salamon2014UrbanSound}
Justin Salamon, Christopher Jacoby, and Juan~Pablo Bello,
\newblock ``A dataset and taxonomy for urban sound research,''
\newblock in {\em ACM MM}, 2014, pp. 1041--1044.

\bibitem{hershey2021benefit}
Shawn Hershey, Daniel~PW Ellis, Eduardo Fonseca, Aren Jansen, Caroline Liu, R~Channing Moore, and Manoj Plakal,
\newblock ``The benefit of temporally-strong labels in audio event classification,''
\newblock in {\em IEEE ICASSP}, 2021.

\bibitem{bilen2020framework}
{\c{C}}a{\u{g}}da{\c{s}} Bilen, Giacomo Ferroni, Francesco Tuveri, Juan Azcarreta, and Sacha Krstulovi{\'c},
\newblock ``A framework for the robust evaluation of sound event detection,''
\newblock in {\em IEEE ICASSP}, 2020.

\bibitem{gromov1987hyperbolic}
Mikhael Gromov,
\newblock ``Hyperbolic groups,''
\newblock in {\em Essays in group theory}. Springer, 1987.

\bibitem{khrulkov2020hyperbolic}
Valentin Khrulkov, Leyla Mirvakhabova, Evgeniya Ustinova, Ivan Oseledets, and Victor Lempitsky,
\newblock ``Hyperbolic image embeddings,''
\newblock in {\em IEEE CVPR}, 2020.

\bibitem{mcinnes2018umap}
Leland McInnes, John Healy, Nathaniel Saul, and Lukas Großberger,
\newblock ``Umap: Uniform manifold approximation and projection,''
\newblock {\em Journal of Open Source Software}, 2018.

\bibitem{smith2023reshaping}
Aaron~Asael Smith, Rui Li, and Zion Tsz~Ho Tse,
\newblock ``Reshaping healthcare with wearable biosensors,''
\newblock {\em Scientific Reports}, vol. 13, no. 1, pp. 4998, 2023.

\bibitem{suris2021learning}
D{\'\i}dac Sur{\'\i}s, Ruoshi Liu, and Carl Vondrick,
\newblock ``Learning the predictability of the future,''
\newblock in {\em CVPR}, 2021.

\end{thebibliography}
